\documentclass[prd,aps,twocolumn,preprintnumbers,showpacs,10pt]{revtex4}
\usepackage{graphicx}
\usepackage{epstopdf}
\usepackage{color}
\usepackage{epsfig}
\def\bold#1{\setbox0=\hbox{$#1$}%
     \kern-.025em\copy0\kern-\wd0
     \kern.05em\%\baselineskip=18ptemptcopy0\kern-\wd0
     \kern-.025em\raise.0433em\box0 }
\def\slash#1{\setbox0=\hbox{$#1$}#1\hskip-\wd0\dimen0=5pt\advance
         to\wd0{\hss\sl/\/\hss}}
\newcommand{\be}{\begin{equation}}
\newcommand{\ee}{\end{equation}}
\newcommand{\bea}{\begin{eqnarray}}
\newcommand{\eea}{\end{eqnarray}}
\newcommand{\nn}{\nonumber}
\newcommand{\dd}{\displaystyle}

\newcommand{\spur}[1]{\not\! #1 \,}
\begin{document}

\preprint{BARI-TH/618-10}
\title{Open charm meson spectroscopy:\\ Where to place the latest piece of the puzzle}
\author{ Pietro Colangelo and Fulvia  De Fazio}
\affiliation{Istituto Nazionale di Fisica Nucleare, Sezione di Bari, Italy}
\begin{abstract}
We discuss how to classify the  $c{\bar s}$ meson $D_{sJ}(3040)$
recently discovered by the  BaBar Collaboration. We consider four
possible assignments, together with  signatures  useful to
distinguish among them.
 \end{abstract}

\maketitle

In the infinite heavy quark mass limit, heavy-light $Q{\bar q}$
mesons can be conveniently classified in doublets labeled by the
value of the total angular momentum ${\vec s}_\ell= {\vec s}_{\bar
q}+{\vec \ell} $ of the light degrees of freedom (light antiquark
and gluons) with respect to the heavy quark  $Q$ \cite{rev0}.
${\vec s}_{\bar q}$ is the spin of $\bar q$ and  ${\vec\ell}$ the
orbital angular momentum. For $\ell=0$ ($s$-wave states in the
constituent quark model), the doublet has $\dd s_\ell=\frac{1}{2}$
and consists of  two states (denoted as $(P,P^*)$,  with $P$  a
generic heavy meson)
 having  spin-parity $J^P_{s_\ell}=(0^-,1^-)_{1/2}$.  $p$-wave states,  with  $\ell=1$, form two doublets: $(P^*_0,P^\prime_1)$ with
$J^P_{s_\ell}=(0^+,1^+)_{1/2}$, and $(P_1,P^*_2)$ with $J^P_{s_\ell}=(1^+,2^+)_{3/2}$.
For $\ell=2$,  $d$-wave states give rise to other two doublets: $(P^*_1,P_2)$ with
$J^P_{s_\ell}=(1^-,2^-)_{3/2}$,  and $(P_2^{\prime *},P_3)$ with $J^P_{s_\ell}=(2^-,3^-)_{5/2}$.

The construction proceeds
further, and has been successfully applied to classify beauty mesons, as well as mesons with  a charmed quark, albeit in the  latter case
the finite heavy quark mass corrections can be  important.

In the case of $c{\bar s}$ mesons, both the lowest lying states,
both the radial excitations (whose members of the various
doublets we  denote by a tilde) can be  described using this
classification scheme. Among the known $c \bar s$ hadrons, the two
mesons $D_s(1969)$ and $D^*_s(2112)$ fill the doublet with
$J^P_{s_\ell}=(0^-,1^-)_{1/2}$, while one can identify
$(D_{s0}^*(2317),D_{s1}^\prime(2460))$ and
$(D_{s1}(2536),D_{s2}^*(2573))$ with the doublets
$J^P_{s_\ell}=(0^+,1^+)_{1/2}$ and $J^P_{s_\ell}=(1^+,2^+)_{3/2}$
respectively (the discovery of $D_{s0}^*(2317)$ and
$D_{s1}^\prime(2460)$ is rather recent \cite{exp}, and their
classification has been the subject of  discussions
\cite{reviews}).

More recently, other candidates have been added to this list:
$D_{sJ}(2860)$, discovered by the BaBar Collaboration
\cite{Aubert:2006mh}, and $D_{sJ}(2710)$, discovered by the Belle and BaBar
Collaborations  \cite{Brodzicka:2007aa,Aubert:2006mh}, both observed  in the  $DK$ final state. In the case of
$D_{sJ}(2710)$, it has been possible to determine the spin-parity:
$J^P=1^-$,  studying its production in $B$ decays. As for $D_{sJ}(2860)$, the  assignments $J^P=3^-$, with radial quantum
number $n=1$ \cite{Colangelo:2006rq}, and $J^P=0^+$, with $n=2$
(i.e. the first radial excitation of $D_{sJ}(2317)$) \cite{zeropiu} have been proposed soon after the discovery.

The latest piece of information comes from the BaBar Collaboration
\cite{Aubert:2009di},  thanks to a new analysis of  $DK$
and $D^*K$ final states.  $D_{sJ}(2710)$ and
$D_{sJ}(2860)$ are seen decaying to both $DK$ and $D^*K$, hence they  have natural parity $J^P=1^-, \, 2^+, \,
3^-, \cdots$. The decay into $D^* K$  excludes the  assignment $J^P=0^+$ for $D_{sJ}(2860)$. Other information comes from the measurement of the  ratios
\bea
{BR(D_{sJ}(2710) \to D^*K) \over BR(D_{sJ}(2710) \to
DK)}=&& \hskip -0.2cm 0.91 \pm 0.13_{stat} \pm 0.12_{syst} \nn \\
{BR(D_{sJ}(2860) \to D^*K) \over BR(D_{sJ}(2860) \to DK)}=&&
\hskip -0.2cm 1.10 \pm 0.15_{stat} \pm 0.19_{syst}  \,\, \nn  \\
\label{exp1}\eea
where $D^{(*)}K$ is the sum over the final states $D^{(*)0}K^+$
and $D^{(*)^+}K_S^0$  \cite{Aubert:2009di}.  Comparing these data with the
predictions in \cite{Colangelo:2007ds}, it seems very likely that
$D_{sJ}(2710)$ is  the first radial excitation of $D_s^*(2112)$,
while   the case of $D_{sJ}(2860)$  requires
further  study  since the measured ratio in  (\ref{exp1}) is larger than the theoretical prediction
$\dd {BR(D_{sJ}(2860) \to D^*K) \over BR(D_{sJ}(2860) \to DK)} \simeq 0.39 \,\,\,\,$  \cite{vb2}.

In the same analysis,  the BaBar Collaboration  observed
 another  broad structure in the $D^* K$ distribution, $D_{sJ}(3040)$, with  \cite{Aubert:2009di}
  \bea
M(D_{sJ}(3040))&=& 3044 \pm 8_{stat}
(^{+30}_{-5})_{syst} \,\,\, {\rm MeV} \nn \\
\Gamma(D_{sJ}(3040)) &=& 239 \pm 35_{stat} (^{+46}_{-42})_{syst}
\,\,\, {\rm MeV} \,\,. \hspace*{0.5cm}
 \eea

Studies of angular distributions for this state have not been
attempted, due to the limited statistics.  Here we discuss possible classifications for this new $c{\bar s}$
meson using the only information available, the measured mass, width and decay mode,
together with the full set of information concerning the other levels/doublets.

%%%%%%%%%%%%%%%%%%%%%%%%%%%%%%%%%%%%%%%%%%%%
\begin{table*}
\centering \caption{ \label{schema} Known $c \bar s$ mesons  organized
according to $s_\ell^P$ and $J^P$; the measured masses  are indicated. Allowed possibilities for $D_{sJ}(3040)$ are  displayed
as ${\bf D_{sJ}(3040)?}$.}
   \begin{ruledtabular}
\begin{tabular}{c  c c c c c }
 $s_\ell^P  $  &$\frac{1}{2}^-$  &   $\frac{1}{2}^+$ &  $\frac{3}{2}^+$  & $\frac{3}{2}^-$ &$\frac{5}{2}^-$ \\ \hline
  $(n=1)$ & &&&& \\
  %    \hline
  $J^P=s_\ell^P-{1\over 2}$&$D_s (1965) \,\, (0^-)  $  & $D_{sJ}(2317) \,\, (0^+)$ &
$D_{s1}(2536) \,\, (1^+) $ & $(1^-)$& ${\bf
D_{sJ}(3040)?}(2^-)$\\
 $J^P=s_\ell^P+{1\over 2}$&$D_s^*(2112) \,\, (1^-)$ & $D_{sJ}(2460)   \,\, (1^+)$
  & $D_{s2}^*(2573) \,\,
  (2^+)  $ &  ${\bf
D_{sJ}(3040)?}(2^-)$&  $D_{sJ}(2860)? (3^-)$\\
  \hline
  $(n=2)$ & &&&& \\
%      \hline
  $J^P=s_\ell^P-{1\over 2}$&  $(0^-)  $& $(0^+)  $ & ${\bf
D_{sJ}(3040)?}(1^+)  $
 &  $(1^-)  $& $(2^-)  $ \\
 $J^P=s_\ell^P+{1\over 2}$& $ D_{sJ}(2710) \,\, (1^-)$ &  ${\bf
D_{sJ}(3040)?}(1^+)  $ & $(2^+)  $   & $(2^-)  $& $(3^-)  $\\
   \end{tabular}
\end{ruledtabular}
\end{table*}
%%%%%%%%%%%%%%%%%%%%%%%%%%%%%%%%%%%%%%%%%%%%
Since $D_{sJ}(3040)$  decays to $D^*K$ and not to $DK$ ,  it has unnatural parity, with possible  $J^P=1^+, \, 2^-, \, 3^+, \cdots$.
The lightest    not yet observed states with such quantum numbers  are the
the two $J^P=2^-$ states belonging to the $\ell=2$ doublets,  with
$\dd s_\ell=\frac{3}{2}$ and  $\dd s_\ell=\frac{5}{2}$:  we denote them as $D_{s2}$ and $D_{s2}^{\prime *}$, respectively  (the case
$J^P=3^+$ corresponds to a  doublet with $\dd s_\ell=\frac{7}{2}$, which is expected to  be  heavier).
In the case of radial excitations, the identification with the states
with $n=2$,  $J^P=1^+$,  and $\dd s_\ell=\frac{1}{2}$ (the meson ${\tilde
D}_{s1}^\prime$) or   $\dd s_\ell=\frac{3}{2}$ (the meson ${\tilde D}_{s1}$)  is
possible.

These four assignments are indicated  in  Table \ref{schema} which reports the classification of
 all the other known $c \bar s$  mesons ($D_{sJ}(2860)$ has  been assigned to the $\dd s_\ell=\frac{5}{2}$
doublet, with a question mark since confirmation is needed).

Some indications about the  masses of these states come from
potential model calculations. For example, in Ref.\cite{Di
Pierro:2001uu} the spectrum of heavy-light mesons is computed in
the framework of a relativistic quark model (RQM), with results:
\bea
M({\tilde D}_{s1})^{(RQM)} &=& 3114 \,\,{\rm MeV} \nn \\
M({\tilde D}_{s1}^\prime)^{(RQM)} &=& 3165 \,\,{\rm MeV} \nn\\
M(D_{s2})^{(RQM)} &=& 2953 \,\,{\rm MeV} \label{rqm} \\
M( D_{s2}^{*\prime})^{(RQM)} &=& 2900 \,\,{\rm MeV} \,\,.\nn \eea
Notice that, if the identification of $D_{sJ}(2860)$ as the
$J^P_{s_\ell}=3^-_{5/2}$ state were experimentally confirmed, this
would disfavor the assignment of $D_{sJ}(3040)$ to its spin
partner $D_{s2}^{*\prime}$ with  $J^P_{s_\ell}=2^-_{5/2}$ , since a mass inversion in a  spin doublet seems unlikely.
 For a similar reason, one would also disfavor the identification of $D_{sJ}(3040)$ with $D_{s2}$,  although in that case the two
mesons  would belong to  different doublets.

The four  classifications for  $D_{sJ}(3040)$ in Table \ref{schema}  can be discussed computing the  allowed strong
decays.  To this purpose, we work in the heavy quark limit  in
which the various spin doublets are described by effective fields:
$H_a$ for $s_\ell^P={1\over2}^-$ ($a=u,d,s$ is a light flavour
index); $S_a$ and $T_a$ for $s_\ell^P={1\over2}^+$ and
$s_\ell^P={3\over2}^+$, respectively; $X_a$ and $X^\prime_a$ for
the doublets  corresponding  to orbital angular momentum $\ell=2$,
i.e.  $s_\ell^P={3\over2}^-$  and $s_\ell^P={5\over2}^-$:
\bea
H_a & =& \frac{1+{\rlap{v}/}}{2}[P_{a\mu}^*\gamma^\mu-P_a\gamma_5]  \label{neg} \nn  \\
S_a &=& \frac{1+{\rlap{v}/}}{2} \left[P_{1a}^{\prime \mu}\gamma_\mu\gamma_5-P_{0a}^*\right]   \nn \\
T_a^\mu &=&\frac{1+{\rlap{v}/}}{2} \Bigg\{ P^{\mu\nu}_{2a}
\gamma_\nu \nn \\ &-&P_{1a\nu} \sqrt{3 \over 2} \gamma_5 \left[ g^{\mu
\nu}-{1 \over 3} \gamma^\nu (\gamma^\mu-v^\mu) \right]
\Bigg\}  \hspace*{1.2cm} \label{pos2} \\
X_a^\mu &=&\frac{1+{\rlap{v}/}}{2} \Bigg\{ P^{*\mu\nu}_{2a}
\gamma_5 \gamma_\nu \nn \\ &-&P^{*\prime}_{1a\nu} \sqrt{3 \over 2}  \left[
g^{\mu \nu}-{1 \over 3} \gamma^\nu (\gamma^\mu-v^\mu) \right]
\Bigg\}   \nn   \\
X_a^{\prime \mu\nu} &=&\frac{1+{\rlap{v}/}}{2} \Bigg\{
P^{\mu\nu\sigma}_{3a} \gamma_\sigma \nn \\ &-&P^{*'\alpha\beta}_{2a}
\sqrt{5 \over 3} \gamma_5 \Bigg[ g^\mu_\alpha g^\nu_\beta - {1
\over 5} \gamma_\alpha g^\nu_\beta (\gamma^\mu-v^\mu) \nn \\ &-&  {1 \over
5} \gamma_\beta g^\mu_\alpha (\gamma^\nu-v^\nu) \Bigg] \Bigg\}
\nn \eea
with the various operators annihilating
mesons of four-velocity $v$ which is conserved in  strong
interaction processes (the heavy field operators  contain a factor
$\sqrt{m_P}$ and have dimension $3/2$).

Let us  consider  decays with the emission of a light
pseudoscalar meson. The octet of light pseudoscalar mesons is
introduced considering the fields:
 $\displaystyle \xi=e^{i {\cal M} \over
f_\pi}$, $\Sigma=\xi^2$,  and  the matrix ${\cal M}$ containing
$\pi, K$ and $\eta$ fields ($f_{\pi}=132 \; $ MeV):
\begin{equation}
{\cal M}= \left(\begin{array}{ccc}
\sqrt{\frac{1}{2}}\pi^0+\sqrt{\frac{1}{6}}\eta & \pi^+ & K^+\nonumber\\
\pi^- & -\sqrt{\frac{1}{2}}\pi^0+\sqrt{\frac{1}{6}}\eta & K^0\\
K^- & {\bar K}^0 &-\sqrt{\frac{2}{3}}\eta
\end{array}\right) \label{pseudo-octet}
\end{equation}

At the leading order in the heavy quark mass and light meson momentum expansion, the decays  $F \to H M$
$(F=H,S,T,X,X^\prime$ and $M$ a light pseudoscalar meson) are
described by the Lagrangian interaction  terms \cite{hqet_chir}:
\bea
{\cal L}_H &=& \,  g \, Tr \Big[{\bar H}_a H_b \gamma_\mu\gamma_5 {\cal A}_{ba}^\mu \Big] \nn \\
{\cal L}_S &=& \,  h \, Tr \Big[{\bar H}_a S_b \gamma_\mu \gamma_5 {\cal A}_{ba}^\mu \Big]\, + \, h.c.  \nn \\
{\cal L}_T &=&  {h^\prime \over \Lambda_\chi}Tr\Big[{\bar H}_a T^\mu_b (i D_\mu {\spur {\cal A}}+i{\spur D} { \cal A}_\mu)_{ba} \gamma_5\Big] + h.c.    \nn \\
{\cal L}_X &=&  {k^\prime \over \Lambda_\chi}Tr\Big[{\bar H}_a X^\mu_b
(i D_\mu {\spur {\cal A}}+i{\spur D} { \cal A}_\mu)_{ba} \gamma_5\Big] + h.c.   \,\,\,\,\,\,\,\,\,\,\,\,  \\
{\cal L}_{X^\prime} &=&  {1 \over {\Lambda_\chi}^2}Tr\Big[{\bar H}_a
X^{\prime \mu \nu}_b \big[k_1 \{D_\mu, D_\nu\} {\cal A}_\lambda \nn \\ &+& k_2
(D_\mu D_\lambda { \cal A}_\nu + D_\nu D_\lambda { \cal
A}_\mu)\big]_{ba}  \gamma^\lambda \gamma_5\Big] + h.c.  \nn
 \label{lag-hprimo} \eea
where ${\cal A}_{\mu
ba}=\frac{i}{2}\left(\xi^\dagger\partial_\mu \xi-\xi
\partial_\mu \xi^\dagger\right)_{ba}$.
$\Lambda_\chi$ is  the chiral symmetry-breaking scale: we use $\Lambda_\chi = 1 \, $ GeV. ${\cal L}_S$ and ${\cal L}_T$ describe
transitions of positive parity heavy mesons with the emission of
light pseudoscalar mesons in $s-$ and $d-$ wave, respectively,  and $g,
h$ and $h^\prime$ are the   effective coupling constants.  On
the other hand, ${\cal L}_X$ and ${\cal L}_{X^\prime}$  describe
the transitions of higher mass mesons of negative parity with the
emission of light pseudoscalar mesons in $p-$ and $f-$ wave,  with
coupling constants $k^\prime$, $k_1$ and $k_2$.

At the same  order in the expansion in the light meson momentum,
the structure of the Lagrangian terms for radial excitations of
the $H$, $S$ and $T$ doublets does not change,  being only
dictated by the spin-flavour and chiral symmetries, but  the
coupling constants $g, h$ and $h^\prime$ must be  replaced
by $\tilde g, \tilde h$ and $\tilde h^\prime$. The advantage of
this formulation is that meson transitions  into final states
obtained  by $SU(3)$ and heavy quark spin rotations can be related
in a straightforward way.

With the effective Lagrangians in Eq.(\ref{lag-hprimo}) we can evaluate the
strong decays of $D_{sJ}(3040)$ to a charmed meson and a light
pseudoscalar one in correspondence to  the four  classifications in Table  \ref{schema}.
In particular, the ratio \be R_1={\Gamma(D_{sJ}(3040) \to D_s^*
\eta) \over \Gamma(D_{sJ}(3040) \to D^* K)} \ee
($D^{(*)}K=D^{(*)0}K^+$ + $D^{(*)^+}K_S^0$) can be computed, with results for the various assignments:
\bea
R_1({\tilde D}_{s1}^\prime)&=&0.34 \nn \\
 R_1({\tilde D}_{s1})&=&0.20 \nn \\
 R_1( D_{s2})&=&0.245 \label{R1} \\
  R_1(D_{s2}^{* \prime})&=&0.143 \,\,. \nn
 \eea
It is important to notice that the dependence on the effective couplings
 cancels in the ratio. The spread among the various predictions  permits to discriminate
among the assignments, in particular between ${\tilde D}_{s1}^\prime$ and $D_{s2}^{* \prime}$.

The mass of  $D_{sJ}(3040)$ is large enough to allow other decay
modes. Decays to the members of the doublets with $s_\ell^P={1
\over 2}^+$ or $s_\ell^P={3 \over 2}^+$ and a pseudoscalar meson
are possible: these are the modes $(D_0^*,D_1^\prime)K$ and
$(D_1,D_2^*)K$. As for channels with the $\eta$ in the final
state, the only  kinematically allowed mode is $D_{s0}^*\eta$ (for
$D_{s1}^\prime \eta$  the available phase space is  tiny). The
features of such decay modes are different for the four considered
assignments. At the leading order in ${1 \over m_c}$ expansion, the states
 ${\tilde D}^\prime_{s1}$ and ${\tilde D}_{s1}$ can
decay to $D^*_0 K$, $D^*_{s0} \eta$, $D^\prime_1K$, $D_1 K$ and
$D^*_2K$ in $p-$ wave;  $D_{s2}$ and $D_{s2}^{*\prime}$ both decay
to $D^*_0 K$, $D^*_{s0} \eta$, $D^\prime_1K$  in $d-$ wave; 
$D_{s2}^{*\prime}$ can decay to $D_1 K$ and $D_2^* K$ in $d-$
wave,
 while in the case of $D_{s2}$
the decay to $D_2^* K$ proceeds in $s-$ wave  and the decay into
$D_1 K$ is allowed  at ${\cal O}\left(1 \over m_c \right)$ in $d-$
wave.

Other  kinematically allowed modes  of  $D_{sJ}(3040)$  are those with the emission of
a  light vector meson, specifically decays into $DK^*$ or $D_s \phi$
\footnote{The decay into $D^*K^*$ is severely  phase
space suppressed.}. It is possible to describe such decay modes using an
approach  based on effective
Lagrangian terms  analogous to the one followed above. In the case of light vector mesons in the final
state, the method has been developed in Ref.\cite{Casalbuoni:1992gi} on the
basis on the hidden gauge symmetry idea \cite{Bando:1985rf}.  We
refer to the review \cite{Casalbuoni:1996pg} for a detailed
description; here we only mention that in this approach the octet
of light vector mesons  is described by the field $\rho_\mu =i
\displaystyle{g_V \over \sqrt{2}} {\hat \rho}_\mu$, where ${\hat
\rho}_\mu$ is a hermitian $3 \times 3$ matrix analogous to
(\ref{pseudo-octet}), containing the light vector meson fields
$\rho^{\pm,0}$, $K^{*\pm}$, $K^{*0}$, ${\bar K}^{*0}$, $\omega_8$.
The  constant $g_V$ can be fixed to the value $g_V=5.8$ by the
Kawarabayashi-Sukuzi, Riazuddin-Fayyazuddin relations
\cite{Kawarabayashi:1966kd}.

In Ref.\cite{Casalbuoni:1996pg} the effective Lagrangian terms describing the transitions $S \to H V$ and $T \to HV$ are
reported, $S$ and $H$ representing the heavy doublets $S_a$ and $H_a$ in Eq. (\ref{pos2}) and $V$ denoting a generic light vector
meson. They read:
\bea {\cal L}_{SHV} &=& i \zeta_S Tr
\Big[{\bar S}_a H_b \gamma_\mu ({\cal V}^\mu -\rho^\mu)_{ba}\Big] \nn \\ &+& i \mu_S Tr \Big[{\bar S}_a H_b \sigma^{\lambda \nu}
F_{\lambda \nu}(\rho)_{ba} \Big] \nn \\
{\cal L}_{THV} &=& i \zeta_T Tr \Big[{\bar H}_a T_b^\mu  ({\cal
V}_\mu -\rho_\mu)_{ba} \Big]  \\ &+& i \mu_T Tr \Big[{\bar H}_a T_b^\mu
v^\nu F_{\mu \nu}(\rho)_{ba} \Big]  \nn\label{lag-rho} \eea
where
${\cal V}_{\mu ba}=\frac{i}{2}\left(\xi^\dagger\partial_\mu
\xi+\xi \partial_\mu \xi^\dagger\right)_{ba}$, and $\zeta_{S,T}$, $\mu_{S,T}$ are effective coupling constants, and
 the field strength tensor $F_{\mu \nu} $ is defined as $F_{\mu \nu}(\rho)=\partial_\mu \rho_\nu -
\partial_\nu \rho_\mu+[\rho_\mu,\rho_\nu]$.

The calculation of all possible decay modes for each one of the four classifications
of  $D_{sJ}(3040)$ would require the values of several coupling constants,  which are unknown. Nevertheless, some conclusions can be drawn.

One can estimate the widths of the decays into $D K^*$ and $D_s \phi$ for the two
$J^P=1^+$ states by the effective Lagrangians in
(\ref{lag-rho}) using the values of $\zeta$ and $\mu$:
$\zeta_S=0.10$ and  $ \mu_S=-0.10 \,\,{\rm GeV}^{-1}$
computed  for the analogous transitions involving
the heavy mesons belonging to the lowest-lying  doublet \cite{Casalbuoni:1996pg}.
This provides us with hints about   the  contribution of such
modes to the total width of $D_{sJ}(3040)$ in the case it is
either ${\tilde D}_{s1}^\prime$ or ${\tilde D}_{s1}$ (in the case
of $D_{s2}$ and $D_{s2}^{*\prime}$ these decays proceed in
$p$-wave and hence they are expected to contribute less significantly):
\bea
\Gamma({\tilde D}_{s1}^\prime \to D K^*) &\simeq& 95 \,\, {\rm MeV}
\nn \\
\Gamma({\tilde D}_{s1}^\prime \to D_s \phi) &\simeq& 44 \,\, {\rm MeV}
\label{DVdecays} \\
\Gamma({\tilde D}_{s1} \to D K^*) &\simeq& 15 \,\, {\rm MeV}
\nn \\
\Gamma({\tilde D}_{s1} \to D_s \phi) &\simeq& 6 \,\, {\rm MeV} \nn
\,\,.
 \eea
It can be noticed that the contribution of these modes to the full width should be
sizable only in the case of ${\tilde D}_{s1}^\prime$;   therefore, this assignment
can be disentangled through the experimental analysis of the $D K^*$ and $D_s \phi$
 decay channels.

%%%%%%%%%%%%%%%%%%%%%%%%%%%%%%%%%%%%%%%%%%%%
\begin{table*}
\centering \caption{Features of the decay modes of $D_{sJ}(3040)$
and of its spin partner at leading order in ${1 \over m_c}$
expansion for the four proposed assignments.}
\begin{ruledtabular}
\begin{tabular}{c |  c | c |c | c }
\small   decay modes & ${\tilde D}_{s1}^\prime$ $(n=2,\,
J^P_{s_\ell}=1^+_{1/2})$ & ${\tilde D}_{s1}$ $(n=2,\,
J^P_{s_\ell}=1^+_{3/2})$ &
 $ D_{s2}$ $(n=1,\, J^P_{s_\ell}=2^-_{3/2})$ & $ D_{s2}^{* \prime}$ $(n=1,\, J^P_{s_\ell}=2^-_{5/2})$ \\
      \hline \hline
$D^* K$, $D^*_s \eta$ & $s-$ wave & $d-$ wave & $p-$ wave & $f-$
wave \\ \hline $R_1$ & 0.34 & 0.20 & 0.245 & 0.143\\ \hline \hline
$D^*_0 K$, $D^*_{s0} \eta$, $D_1^\prime K$ & $p-$ wave & $p-$ wave
& $d-$ wave & $d-$ wave \\ \hline \hline $D_1 K$ & $p-$ wave &
$p-$ wave & - & $d-$ wave \\ \hline $D_2^* K$ & $p-$ wave & $p-$
wave & $s-$ wave & $d-$ wave \\ \hline \hline $D K^*$, $D_s
\phi$ & $s-$ wave & $s-$ wave & $p-$ wave & $p-$ wave \\
\cline{2-5} &$\Gamma\simeq 140$
MeV & $\Gamma\simeq 20$ MeV  & negligible & negligible \\
\hline \hline
  & \multicolumn{4}{c }  {\rm spin partner} \\ \hline \hline
   &${\tilde D}_{s0}^*$ $(n=2,\, J^P_{s_\ell}=0^+_{1/2})$ & ${\tilde
D}_{s2}^*$ $(n=2,\, J^P_{s_\ell}=2^+_{3/2})$ & $D_{s1}^*$ $(n=1,\,
J^P_{s_\ell}=1^-_{3/2})$ & $D_{s3}$ $(n=1,\,
J^P_{s_\ell}=3^-_{5/2})$\\ \hline \hline
$DK$, $D_s \eta$ & $s-$
wave & $d-$ wave & $p-$ wave & $f-$ wave
\\ \hline
$D^*K$, $D_s^* \eta$ & - & $d-$ wave & $p-$ wave & $f-$ wave
\\ \hline \hline
$D^*_0 K$, $D^*_{s0} \eta$ & - & - & $d-$ wave & - \\ \hline
$D_1^\prime K$ & $p-$ wave & $p-$ wave & $d-$ wave & $d-$ wave
\\ \hline \hline
$D_1 K$ & $p-$ wave & $p-$ wave & $s-$ wave & $d-$ wave \\ \hline
$D_2^* K$ & - & $p-$ wave & - & $d-$ wave \\

   \end{tabular}
\end{ruledtabular}\label{summary}
\end{table*}
%%%%%%%%%%%%%%%%%%%%%%%%%%%%%%%%%%%%%%%%%%%%

After having discussed   four  possible classifications for $D_{sJ}(3040)$,
it is interesting to consider some features of  its spin partner
 in the various cases  \footnote{The allowed decay modes are
listed assigning the mass  $M=3040$ MeV  to the spin partner. For a
slightly different mass, some channels could no more be
allowed. For example, some of the modes listed here for the
$J^P=3^-$ state have not been considered in  \cite{Colangelo:2006rq},  since such a state has a lower
mass, $M=2860$ MeV.}.
\begin{itemize}
\item If   $D_{sJ}(3040)$ is ${\tilde D}_{s1}^\prime$
($s_\ell^P=\frac{1}{2}^+$, $J^P=1^+$, $n=2$), its spin partner  is
${\tilde D}_{s0}^*$,  a $J^P=0^+$ state,   the first radial
excitation of $D_{sJ}(2317)$. This state can decay to $DK$ and
$D_s\eta$ in $s$-wave;  $p$-wave decays to $D_1^\prime K$ and $D_1
K$ are also allowed.
\item  If   $D_{sJ}(3040)$ is
${\tilde D}_{s1}$ ($s_\ell^P=\frac{3}{2}^+$, $J^P=1^+$, $n=2$),
its spin partner is ${\tilde D}_{s2}^*$ with $J^P=2^+$. It is allowed to decay to
$DK$, $D_s\eta$, $D^*K$, $D_s^*\eta$, $DK^*$ and $D_s \phi$ in
$d$-wave,  and to $D_1^\prime K$,  $D_1 K$ and $D_2^* K$ in $p$-wave.
\item  If   $D_{sJ}(3040)$ is $ D_{s2}$ ($s_\ell^P=\frac{3}{2}^-$, $J^P=2^-$, $n=1$),
 its spin partner is the vector meson $D_{s1}^*$ with $J^P=1^-$. It can
decay to $DK$, $D_s\eta$, $D^*K$, $D_s^*\eta$, $DK^*$ and $D_s
\phi$ in $p$-wave,  to $D_0^*K$, $D_{s0}^* \eta$ and $D_1^\prime
K$   in $d-$ wave and to $D_1 K$ in $s-$ wave. The decay to $D_2^*
K$ is allowed  at ${\cal O}\left(1 \over m_c \right)$ in $d-$
wave.
\item
  If   $D_{sJ}(3040)$ is  $ D_{s2}^{*\prime}$ ($s_\ell^P=\frac{5}{2}^-$, $J^P=2^-$, $n=1$),
   its spin partner  is $D_{s3}$ with $J^P=3^-$, decaying to $DK$, $D_s\eta$, $D^*K$ and $D_s^*\eta$,
$DK^*$, $D_s \phi$ in $f-$ wave, and to $D_1^\prime K$, $D_1 K$
and $D_2^* K$ in $d-$ wave.
\end{itemize}
%%%%%%%%%%%%%%%%%%%%%%%%%%%%%%%%%%%%%%%%%%
\begin{figure}[b]
\includegraphics[width=0.4\textwidth]{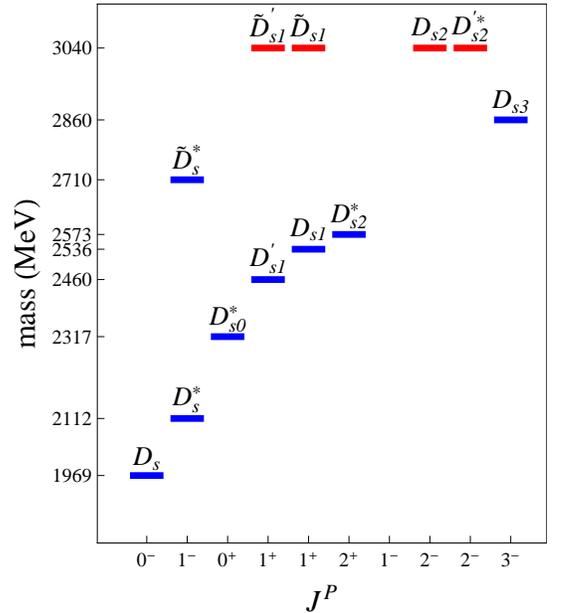}
\caption{Spectrum of the $c \bar s$ system. All observed $D_{sJ}$ states, with mass indicated on the $y$ axis,  are assigned to a level with
 $J^P$ and proper name.
The four  assignments discussed for $D_{sJ}(3040)$ are shown  in correspondence to the mass value $M=3040$ MeV.}
\label{cspectrum}
\end{figure}
%%%%%%%%%%%%%%%%%%%%%%%%%%%%%%%%%%%%%%%%%%

Since $D_{sJ}(3040)$ has a  broad width, we expect that also its
spin partner shares the same feature. Considering the previous
list, we  can argue that ${\tilde D}_{s0}^*$ is  broad due to its
 $s$-wave decays into $DK$ and $D_s \eta$. Also
$D_{s1}^*$ has allowed $s$-wave decays, but only to  $D_1 K$ which
is suppressed by phase space effects.

The identification of $D_{sJ}(3040)$ with ${\tilde D}_{s1}^\prime$ is
supported in Refs.\cite{Chen:2009zt,Sun:2009tg} on the basis of the  $c{\bar s}$
  mass spectrum  \cite{Chen:2009zt}
  or of the decay widths computed  in the $^3P_0$ model  \cite{Sun:2009tg}.
In the second case,   the identification with
${\tilde D}_{s1}^\prime$ and ${\tilde D}_{s1}$ is discussed: the full widths of
 these two states are computed and compared to the experimental measurement
of $\Gamma(D_{sJ}(3040))$, concluding that for ${\tilde D}_{s1}^\prime$
 the experimental width can be reproduced, with the prediction:
$\displaystyle{\Gamma(D_{sJ}(3040) \to DK^* +D_s \phi) \over
\Gamma(D_{sJ}(3040) \to D^*K +D^*_s \eta)}\simeq 0.79$. In our discussion   we have examined   other information that can be exploited.

To summarize the results of our considerations and compare  the
distinctive features of the four proposed assignments for
$D_{sJ}(3040)$ (illustrated in Fig.\ref{cspectrum} in which all the
known $c \bar s$ states have  been included),  we collect our
findings and observations in Table \ref{summary}. In particular,
we emphasize the role of the final states $D K^*$ and $D_s \phi$,
which  deserve an  experimental investigation.  Moreover, although
the identification with a $J^P=2^-$  $\displaystyle
s_\ell^P=\frac{3}{2}^-$ state seems less probable, this assignment
can be discarded/confirmed studying  the $D_2^* K$ $s-$wave final
state. Search of  the spin partner in each doublet would provide
further information, enriching the $c \bar s$ spectrum recently
disclosed by experiments.

\vspace*{1cm}
\noindent {\bf Acknowledgments} \\
\noindent We thank  A. Palano for discussions. This work was
supported in part by the EU contract No. MRTN-CT-2006-035482,
"FLAVIAnet".

%*****************
% \clearpage
 

\begin{thebibliography}{99}

\bibitem{rev0}
For  reviews see: M.~Neubert,
  %``Heavy quark symmetry,''
  Phys.\ Rept.\  {\bf 245}, 259 (1994);
  %%CITATION = PRPLC,245,259;%%
  F.~De Fazio,
  %``Weak decays of heavy quarks,''
 in {\it At the
Frontier of Particle Physics/Handbook of QCD}, ed. by M. Shifman
(World Scientific, Singapore, 2001),  page  1671,
arXiv:hep-ph/0010007.
  %%CITATION = HEP-PH/0010007;%%

 \bibitem{exp}
B.~Aubert {\it et al.}  [BABAR Collaboration],
%``Observation of a narrow meson decaying to D/s+ pi0 at a mass of
%2.32-GeV/c**2,''
 Phys. Rev. Lett.  {\bf 90}, 242001 (2003);
%[arXiv:hep-ex/0304021].
%%CITATION = HEP-EX 0304021;%%
Y.~Mikami {\it et al.} [Belle Collaboration], Phys. Rev. Lett. {\bf 92}, 012002 (2004);
%%CITATION = HEP-EX 0307052;%%
D.~Besson {\it et al.}  [CLEO Collaboration], Phys.\ Rev.\ D {  \bf 68}, (2003) 032002;
%[arXiv:hep-ex/0305100].
%%CITATION = HEP-EX 0305100;%%
L.~Benussi  [FOCUS Collaboration],
  %``Measurements of masses and widths of excited charm mesons D2*, evidence for
  %broad states and observation of D/s(2317)+ from FOCUS experiment,''
  Int.\ J.\ Mod.\ Phys.\  A {\bf  20}, 549 (2005).
  %%CITATION = IMPAE,A20,549;%%

\bibitem{reviews}
For reviews see: P.~Colangelo, F.~De Fazio and R.~Ferrandes,
  %``Excited charmed mesons: Observations, analyses and puzzles,''
  Mod.\ Phys.\ Lett.\  A {\bf 19},  2083 (2004);
  %%CITATION = MPLAE,A19,2083;%%
  E.~S.~Swanson,
  %``The new heavy mesons: A status report,''
  Phys.\ Rept.\  {\bf 429},  243 (2006);
  %%CITATION = PRPLC,429,243;%%
P.~Colangelo, F.~De Fazio, R.~Ferrandes and S.~Nicotri,
  %``Puzzles in charm spectroscopy,''
  Prog.\ Theor.\ Phys.\ Suppl.\  {\bf 168}, 202 (2007).
  %%CITATION = PTPSA,168,202;%%

\bibitem{Aubert:2006mh}
B.~Aubert {\it et al.}  [BABAR Collaboration],
  %``Observation of a new D/s meson decaying to D K at a mass of
  %2.86-GeV/c**2,''
  Phys. Rev. Lett.  { \bf 97}, 222001 (2006).
  %%CITATION = PRLTA,97,222001;%%

\bibitem{Brodzicka:2007aa}
  J.~Brodzicka {\it et al.}  [Belle Collaboration],
  %``Observation of a new D_sJ meson in B+->D0BD0K+ decays,''
  Phys.\ Rev.\ Lett.\  {\bf 100}, 092001 (2008).
%  [arXiv:0707.3491 [hep-ex]].
  %%CITATION = PRLTA,100,092001;%%

\bibitem{Colangelo:2006rq}
  P.~Colangelo, F.~De Fazio and S.~Nicotri,
  %``$D_{sJ}(2860)$ resonance and the $s_\ell^P={5\over 2}^-$ $c \bar s$ ($c
  %\bar q$) doublet,''
  Phys. Lett.  B {\bf 642}, 48 (2006).
  %%CITATION = PHLTA,B642,48;%%

\bibitem{zeropiu}
  E.~van Beveren and G.~Rupp,
  %``New BABAR state D/sJ(2860) as the first radial excitation of the
  %D/s0*(2317),''
  Phys. Rev. Lett.   {\bf 97}, 202001 (2006);
  %%CITATION = PRLTA,97,202001;%%
 F.~E.~Close, C.~E.~Thomas, O.~Lakhina and E.~S.~Swanson,
  %``Canonical Interpretation of the D/sJ(2860) and D/sJ(2690),''
  Phys. Lett. B {\bf 647}, 159 (2007).
  %%CITATION = PHLTA,B647,159;%%

\bibitem{Aubert:2009di}
 B.~Aubert {\it et al.}  [BABAR Collaboration],
  %``Study of $D_{sJ}$ decays to $D^*K$ in inclusive $e^+e^-$ interactions,''
  Phys.\ Rev.\  D {\bf 80}, 092003 (2009).
%  [arXiv:0908.0806 [hep-ex]].
  %%CITATION = PHRVA,D80,092003;%%

\bibitem{Colangelo:2007ds}
  P.~Colangelo, F.~De Fazio, S.~Nicotri and M.~Rizzi,
  %``Identifying $D_{sJ}(2700)$ through its decay modes,''
  Phys.\ Rev.\  D {\bf 77},  014012 (2008).
 % [arXiv:0710.3068 [hep-ph]].
  %%CITATION = PHRVA,D77,014012;%%

\bibitem{vb2}
Among the various possibilities, one has to investigate the size of $1/m_c$ corrections to this ratio,
waiting for a higher statistics confirmation of the measurement. As an alternative,
the possibility that  $D_{sJ}(2860)$ is a mixing of
$J^P=0^+$ and $J^P=2^+$ components has been suggested by  E.~van Beveren
and G.~Rupp,
  %``Comment on 'Study of D(sJ) decays to D(*)K in inclusive e(+)e(-)
  %interactions',''
  arXiv:0908.1142 [hep-ph].
  %%CITATION = ARXIV:0908.1142;%%

\bibitem{Di Pierro:2001uu}
  M.~Di Pierro and E.~Eichten,
  %``Excited heavy-light systems and hadronic transitions,''
  Phys.\ Rev.\  D {\bf 64}, 114004 (2001).
%  [arXiv:hep-ph/0104208].
  %%CITATION = PHRVA,D64,114004;%%

  \bibitem{hqet_chir}
M.B. Wise, Phys. Rev. D {\bf 45}, R2188   (1992); G. Burdman and J.F.
Donoghue, Phys. Lett. B {\bf 280}, 287  (1992); P. Cho, Phys. Lett. B 285,
145 (1992); H.-Y. Cheng {\it et al.,}  Phys. Rev.  D {\bf 46},  1148 (1992).

\bibitem{Casalbuoni:1992gi}
  R.~Casalbuoni, A.~Deandrea, N.~Di Bartolomeo, R.~Gatto, F.~Feruglio and G.~Nardulli,
  %``Light vector resonances in the effective chiral Lagrangian for heavy
  %mesons,''
  Phys.\ Lett.\  B {\bf 292}, 371 (1992).
%  [arXiv:hep-ph/9209248].
  %%CITATION = PHLTA,B292,371;%%

\bibitem{Bando:1985rf}
  M.~Bando, T.~Kugo and K.~Yamawaki,
  %``On The Vector Mesons As Dynamical Gauge Bosons Of Hidden Local
  %Symmetries,''
  Nucl.\ Phys.\  B {\bf 259}, 493 (1985);
  %%CITATION = NUPHA,B259,493;%%
Phys.\ Rept.\  {\bf 164}, 217 (1988).
  %%CITATION = PRPLC,164,217;%%

\bibitem{Casalbuoni:1996pg}
  R.~Casalbuoni, A.~Deandrea, N.~Di Bartolomeo, R.~Gatto, F.~Feruglio and G.~Nardulli,
  %``Phenomenology of heavy meson chiral Lagrangians,''
  Phys.\ Rept.\  {\bf 281}, 145 (1997).
%  [arXiv:hep-ph/9605342].
  %%CITATION = PRPLC,281,145;%%

\bibitem{Kawarabayashi:1966kd}
  K.~Kawarabayashi and M.~Suzuki,
  %``Partially conserved axial vector current and the decays of vector mesons,''
  Phys.\ Rev.\ Lett.\  {\bf 16}, 255 (1966);
  %%CITATION = PRLTA,16,255;%%
Riazuddin and Fayyazuddin,
  %``Algebra of current components and decay widths of rho and K* mesons,''
  Phys.\ Rev.\  {\bf 147}, 1071 (1966).
  %%CITATION = PHRVA,147,1071;%%

\bibitem{Chen:2009zt}
  B.~Chen, D.~X.~Wang and A.~Zhang,
  %``Interpretation of $D_{sJ}(2632)^+$, $D_{s1}(2700)^\pm$,
  %$D^\star_{sJ}(2860)^+$ and $D_{sJ}(3040)^+$,''
  Phys.\ Rev.\  D {\bf 80}, 071502 (2009).
%  [arXiv:0908.3261 [hep-ph]].
  %%CITATION = PHRVA,D80,071502;%%

\bibitem{Sun:2009tg}
   Z.~F.~Sun and X.~Liu,
  %``Newly observed $D_{sJ}(3040)$ and the radial excitations of P-wave
  %charmed-strange mesons,''
  Phys.\ Rev.\  D {\bf 80}, 074037 (2009).
%  [arXiv:0909.1658 [hep-ph]].
  %%CITATION = PHRVA,D80,074037;%%

 \end{thebibliography}
\end{document}